\documentclass[
 prd,
 amsmath,
 amssymb,
 reprint,
 twocolumn,
]{revtex4-1}

\usepackage{graphicx}
\usepackage{dcolumn}
\usepackage{bm}
\usepackage{braket}
\usepackage{color}
\usepackage{xcolor}
\usepackage{tabularx}
\usepackage{mathrsfs}
\usepackage{hyperref}
\usepackage{caption}
\usepackage{afterpage}
\usepackage{subcaption}

\newcommand{\chirp}{\,\text{M}_\odot}
\hypersetup{
    colorlinks=true,
    linkcolor=blue,
    filecolor=magenta,      
    urlcolor=cyan,
    pdftitle={Overleaf Example},
    pdfpagemode=FullScreen,
    }

\begin{document}

\title{Phase consistency test to identify type II strongly lensed gravitational wave signals using a single event} 

\author{Kelsie~Taylor$^{1,2}$, Derek~Davis$^2$, and Rico~K.~L.~Lo$^{3}$}

\affiliation{$^1$Weinberg Institute for Theoretical Physics, University of Texas at Austin, Austin, TX 78712, USA}
\affiliation{$^2$LIGO, California Institute of Technology, Pasadena, CA 91125, USA}
\affiliation{$^3$Center of Gravity, Niels Bohr Institute, Blegdamsvej 17, 2100 Copenhagen, Denmark}

\date{\today} 

\begin{abstract}
For gravitationally lensed type II signals, the phase of the dominant (2, 2) mode and the higher order (3, 3) mode is offset by $-\pi/12$, or roughly -0.26 radians. Using this, we develop a test for type II imagery by allowing the phases of the (2,2) and (3,3) modes to vary separately and introducing a new waveform parameter to represent the phase offset between the two. We use simulated, asymmetric mass ratio, precessing signals to show that the test can reproduce the $-\pi/12$ phase offset when detected by three detectors for H-L optimal SNR $\gtrsim$ 40 and $\mathcal{M} \leq 30$. We analyze GW190412 and GW190814 using this parameterization, measuring the offset to be $0.13^{+0.22}_{-0.17}$ for GW190412 and $-0.05^{+0.20}_{-0.22}$ for GW190814. We also measure the Bayes factor in support of zero phase offset, $\log_{10} \mathcal{B}_{\Delta \varphi = 0}$, to be $-0.14$ for GW190412 and $0.21$ for GW190814. This implies our results are not strong enough to confidently argue if either event is a type II image, and is consistent with our statistical analysis. 

\end{abstract}

\maketitle 

\section{Introduction}

Gravitational waves (GWs) are perturbations in spacetime that were predicted by Einstein's Theory of General Relativity (GR) in 1916~\cite{Einstein:1916}. In September 2015, the Advanced Laser Interferometer Gravitational-Wave Observatory (LIGO) recorded the first observation of gravitational waves from a binary black hole merger (BBH)~\citep{GW150914}. Since then, the LIGO-Virgo-KAGRA collaboration has announced 90 significant gravitational-wave events from the three observing runs~\cite{GWTC-1, GWTC-2, GWTC-2.1, GWTC-3}. 

As the sensitivity of ground-based gravitational-wave interferometers steadily improves, it is possible to measure increasingly subtle effects in the waveforms of observed signals. 
One such opportunity is testing for the existence of gravitationally lensed GWs: gravitational waves that have been deflected after propagating near massive objects~\cite{1992grle.book, Takahashi:2003ix}. 
In this paper, we focus on strong lensing in geometric optics, which forms multiple images with differing magnifications from a single source. These images can be further categorized into three types based on their solution to the lens equation: type I images consisting of local minimum solutions, type II images consisting of saddle point solutions, and type III images consisting of local maximum solutions~\cite{Blandford:1986zz, Ezquiaga:2020gdt,Ezquiaga:2023xfe, Wang:2021kzt,Vijaykumar:2022dlp}. 
Each solution induces a different phase shift to the observed gravitational wave, but the phase shifts in type I and type III images are degenerate with the phase of a non-lensed waveform. Type II images, however, have a non-degenerate phase shift, making these solutions particularly interesting.

This feature has been previously explored as a way to identify a lensed signal with a single image~\cite{Ezquiaga:2020gdt,Wang:2021kzt, Vijaykumar:2022dlp}.
A wide variety of different investigations of evidence of lensing in recent gravitational-wave catalogs have been conducted~\cite{Hannuksela:2019kle, gwtc2_lensing, Dai:2020tpj, gwtc3_lensing, Janquart:2023mvf}; 
although multiple candidates have been investigated~\cite{Dai:2020tpj,Janquart:2023mvf},
 no gravitational-wave signal has been confidently identified as lensed.

In this work, we explore the measurability of the phase shift expected to be induced in gravitationally lensed type II images from compact binary mergers. 
For a type II image, there is expected to be an observable $-\pi/12$ offset between the (2,2) mode and the (3,3) mode~\cite{Wang:2021kzt}. 
This knowledge can be used to test if a single gravitational-wave signal is a type II image, without requiring the identification of the lens or a second image of the same compact binary merger. 
To do so, we break down our gravitational wave signal into a superposition of spherical harmonics, examining only the dominant $(l, m) = (2, 2)$ mode and subdominant $(3,3)$ mode. (We make the assumption that the (2, 2) mode is dominant and the (3, 3) mode is subdominant, as this is true for quasi-circular compact binary mergers with a decently symmetric mass ratio.) 
We then allow the phases to vary separately, adding a new ``phase offset'' parameter $\Delta \varphi$ to look for the offset. This model was highly inspired by \emph{Capano and Nitz}, who used a model varying masses, spins, and phase parameters between the (2, 2) and (3, 3) modes to test the consistency in GR~\citep{Spectroscopy}, though we only allow the phase to vary and consider lensing instead of GR consistency.

To confirm the validity of this test, we perform Bayesian parameter estimation of our model on a population of simulated signals. In our test, we insert signals with $\Delta \varphi = -\pi/12$ to see what conditions are needed to discover the phase offset through Bayesian sampling. We do this for various sampling techniques, chirp masses, and SNRs to get a comprehensive picture of what is needed to confidently measure this offset. In addition to simulated signals, we also analyze events GW190412~\cite{GW190412} and GW190814~\cite{GW190814} for gravitational lensing, as they are the most likely to have measurable differences between the (2, 2) and (3, 3) modes due to high SNR, asymmetric mass ratio and specified localization areas.
Our analysis of GW190814 is the first to highlight the potential of this event for identifying type II  images.

We discuss these results as follows. In Section \ref{sec:model}, we discuss the specifics of our waveform that we used to generate and sample signals. In Section \ref{sec:sims}, we discuss the results of our simulated tests. In Section \ref{sec:events}, we discuss the results of our tests on events GW190412 and GW190814. Finally, in Section \ref{sec:compare}, we compare our analysis to other results in the literature, and in Section \ref{sec:conclusions}, we discuss general conclusions and future work.  

\section{Model Description} \label{sec:model}

For a gravitational-wave signal associated with a compact binary merger without lensing, we define the signal's source frame such that the $z$-axis is aligned with the direction of the binary's orbital angular momentum at a given reference time. We define the inclination $\iota$ as the angle between the observer's line of sight and the $z$-axis of the source's center of mass frame, and $\phi$ is the azimuthal angle of the observer with respect to this frame. We also denote the direction of the observer from the source to be $-\hat{n}$, allowing us to define the $+$ and $\times$ polarizations of gravitational waves propagating in this direction such that $+$ is in the $\vec{e}_\theta$ direction.

Using this coordinate system, we can express the $+$ and $\times$ polarizations of a gravitational wave signal $h$ in terms of merger parameters. In our case, we choose to do this via decomposing the signal into a spin-weighted spherical harmonic basis with spin weight $-2$. This decomposition for a signal $h$ at a luminosity distance $D_L$ from the source is given by
\begin{equation} 
\begin{split}
& h_+ - i h_\times = \\
& \frac{1}{D_L}\sum_{l = 2}^{\infty} \sum_{m = -l}^l \,_{-2}Y_{lm}(\iota, \phi)A_{lm}(\vec{\lambda})e^{ i (\Psi_{lm}(\vec{\lambda}) + m \varphi)},
\end{split}
\label{decomp}
\end{equation}

where $\,_{-2}Y_{lm}$ is the spin-weighted spherical harmonic, $A_{lm}$ is a function determining the amplitude of the $(l, m)$ mode of the signal and $\Psi_{lm}$ is a function determining the phase of the $(l, m)$ mode of the signal. $\vec{\lambda}$ represents the intrinsic parameters of the merger: the masses $m_1, m_2$ and spins $\vec{\chi}_1, \vec{\chi}_2$. We work in the frequency domain, defining the reference frequency to be 20 Hz and thus the reference time as the time where the dominant mode is at 20 Hz; it is at this time that the phase $\varphi$ is measured.

The predominant term in this sum is given by the $(2,2)$ mode, the dominant quadrupole mode of the wave's signal. This quadrupole mode emits frequencies twice that of the orbital frequency, while the higher order modes emit at higher integer multiples of the orbital frequency~\cite{HM_1,HM_2}. As previously stated, we assume the higher order amplitudes are weaker than the dominant amplitudes, making them harder to measure. For this reason, we only include the dominant $(2,2)$ mode and the subdominant $(3,3)$ mode in the waveforms that we will be using, since we expect these modes to have the highest SNRs. In general, this formalism can be extended to an arbitrary number of modes.  

For strongly lensed type II images, however, there is a slight difference in this decomposition than that in Equation \ref{decomp}. This happens because a type II signal experiences phase shifts from its non-lensed counterpart. In these shifts, the amount the phase of each mode changes compared to the non-lensed signal varies with the value of $|m|$. Thus, each mode of a type II image has its own phase $\varphi_{m}$ that varies between modes. This suggests that we can test for lensing by varying $\varphi$ between modes (l, m), denoting each individual phase as $\varphi_{lm}$. In our model, we vary $\varphi_{22}$ and $\varphi_{33}$ independently of each other to calculate a new parameter, the phase offset: \begin{equation} 
    \Delta \varphi = \varphi_{33} - \varphi_{22}.
    \label{offset}
\end{equation}
For a non-lensed signal, $\Delta \varphi$ is just 0, because the phase stays the same between modes. For a type II signal, however, lensing shifts the phase of the $(l, \pm 2)$ modes by $\pi/4$ from the non-lensed signal, giving the signal a degeneracy of $\pi/2$, while for the $(l, \pm 3)$ modes it shifts the phase by $\pi/6$, giving the signal a degeneracy of $\pi/3$. ~\cite{Ezquiaga:2020gdt, Wang_2021}. Thus, the total offset between the two modes should be $\Delta \varphi = \pi/6-\pi/4 = -\pi/12$, with a degeneracy every $|\pi/3 - \pi/2| = \pi/6$ radians to account for the degeneracy in the individual phase parameters. 
This parameterization has previously been found to faithfully model type II lensed images from binaries with little to no precession and a reasonable approximation for signals from highly precessing binaries~\cite{Ezquiaga:2020gdt}. 

This model thus provides a straightforward way to test for type II images from gravitational lensing. We perform Bayesian parameter estimation using a model containing the $(2,2)$ and $(3,3)$ modes, allowing the phases to vary separately. Thus, in addition to estimating the typical compact binary merger parameters, we also estimate $\Delta \varphi$, to see if the signal shows this $-\pi/12$ offset. We allow $\Delta \varphi$ to vary uniformly between $-\pi/6$ and $\pi/6$ to ignore most degeneracies. For the rest of the parameters, we use typical astrophysical priors (see Appendix~\ref{app:details}).

The waveform we used throughout this work is \texttt{IMRPhenomXPHM}~\cite{Pratten:2020ceb}, which includes both precession effects and higher-order mode effects. We only include the (2,2) and (3,3) modes, and modify the waveform with a mode-dependent phase as explained above. To create the modified waveform, perform our parameter estimation, and create simulated signals, we used the \texttt{Bilby} software package~\cite{Bilby}. We used nested sampling through the sampler \texttt{Dynesty}~\cite{Speagle:2019ivv} to estimate the source parameters.

\section{Simulated Signals} \label{sec:sims}

To test the ability of our model to recover offsets between the (2,2) mode and (3,3) mode, we inject \texttt{IMRPhenomXPHM} waveforms with a known phase offset into colored Gaussian noise with spectra representative of detector data from the LIGO~\cite{aLIGO:2020wna} and Virgo~\cite{Virgo:2022ysc} detectors during the third observing run to simulate a type II image. We use the same realization of the noise for all injections for simplicity. 

We focus on three parameters that impact the measurability of $\Delta \varphi$ -- the chirp mass $\mathcal{M}$, signal SNR (which we vary via the luminosity distance $D_L$), and the number of detectors used to analyze each signal. For all quoted SNRs, we use the H-L optimal SNR, defined as the sum in quadrature of the Hanford detector's SNR and the Livingston detector's SNR after matched filtering. We quote this SNR even in the cases in which we include Virgo, for the purposes of direct comparison.

Our injection investigations can be split into two different SNR regimes: extremely high SNRs (SNR $\gtrsim$ 100)and ``typical'' SNRs (10 $\lesssim$ SNR $\lesssim$ 50). We examined 18 injections for high SNRs and 15 injections for the typical SNRs. Additional analysis details for the results in this section can be found in Appendix~\ref{app:details}.

\subsection{General Results}

\begin{figure*}[tb]
    \centering
    \includegraphics[width=0.44\textwidth]{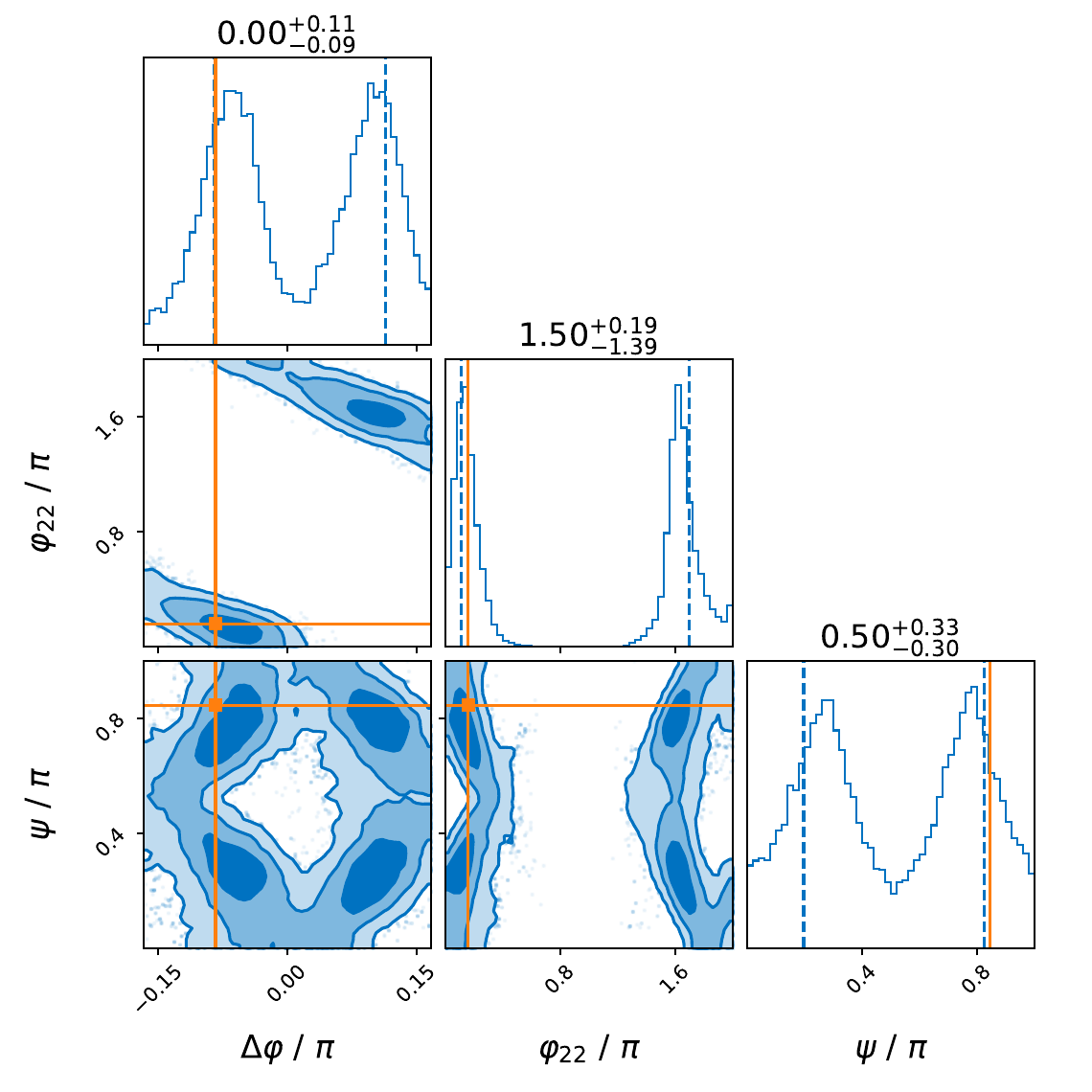}
    \includegraphics[trim={0 1cm 0 3cm},clip,width=0.54\textwidth]{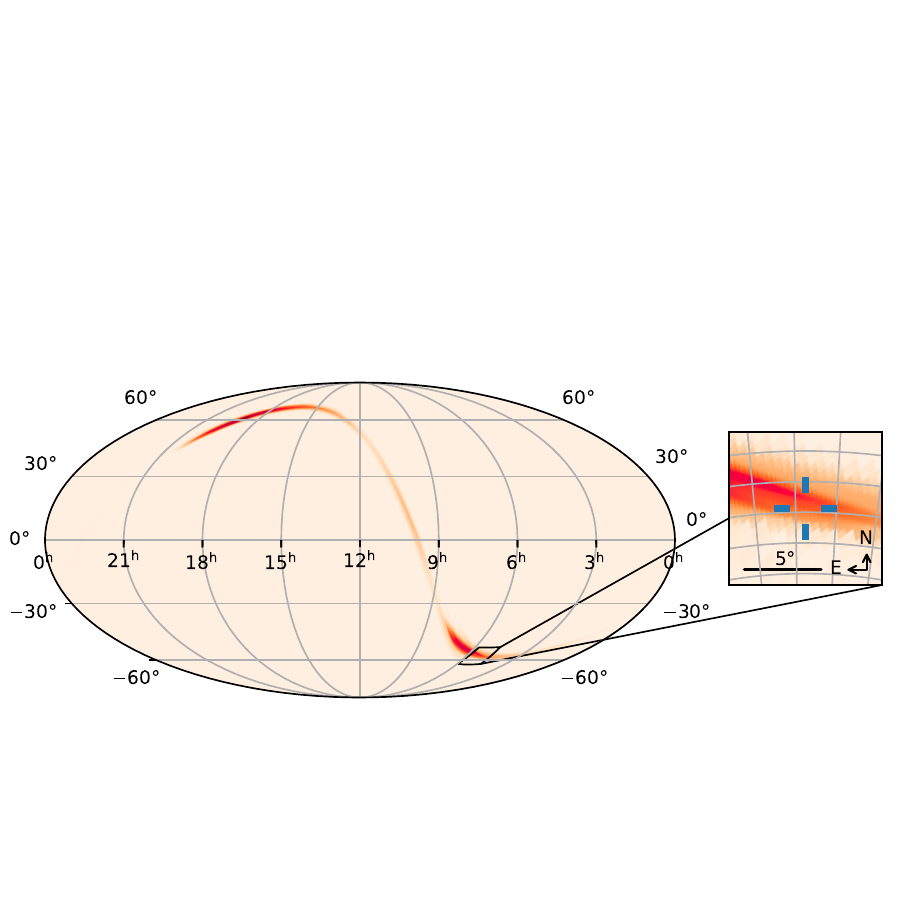}
    \caption{Example of analyses of a simulated signal with network SNR $\sim$50 that is observed only in the LIGO Hanford and LIGO Livingston detectors. Left: corner plot showing the posterior for the phasing parameters of the signal, including the phase-offset used in this work. Right: Inferred skymap for this simulated signal. Cross hairs in each panel indicate injected values. }
    \label{fig:high_snr_2_det}
\end{figure*}

\begin{figure*}[tb]
    \centering
    \includegraphics[width=0.44\textwidth]{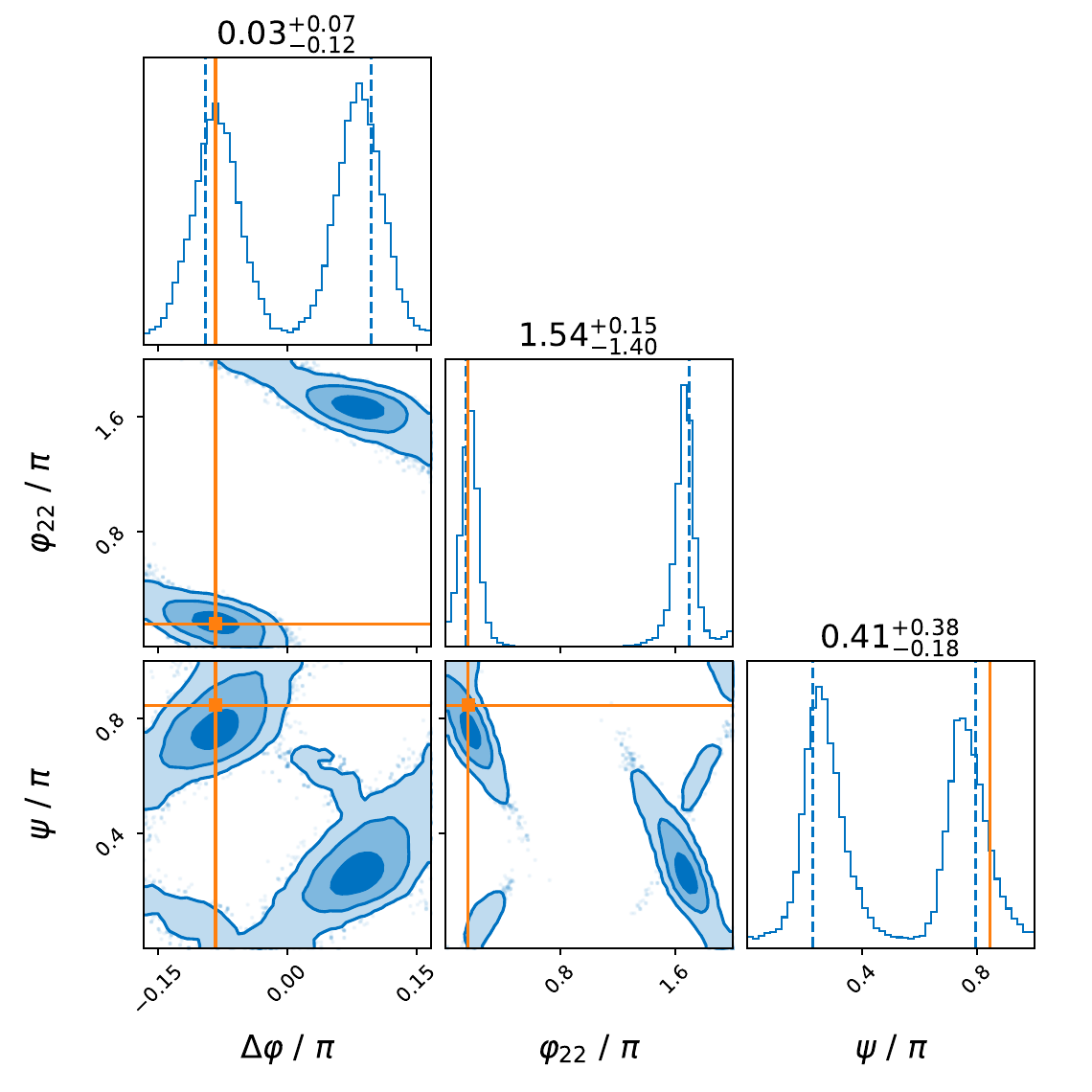}
    \includegraphics[trim={0 1cm 0 3cm},clip,width=0.54\textwidth]{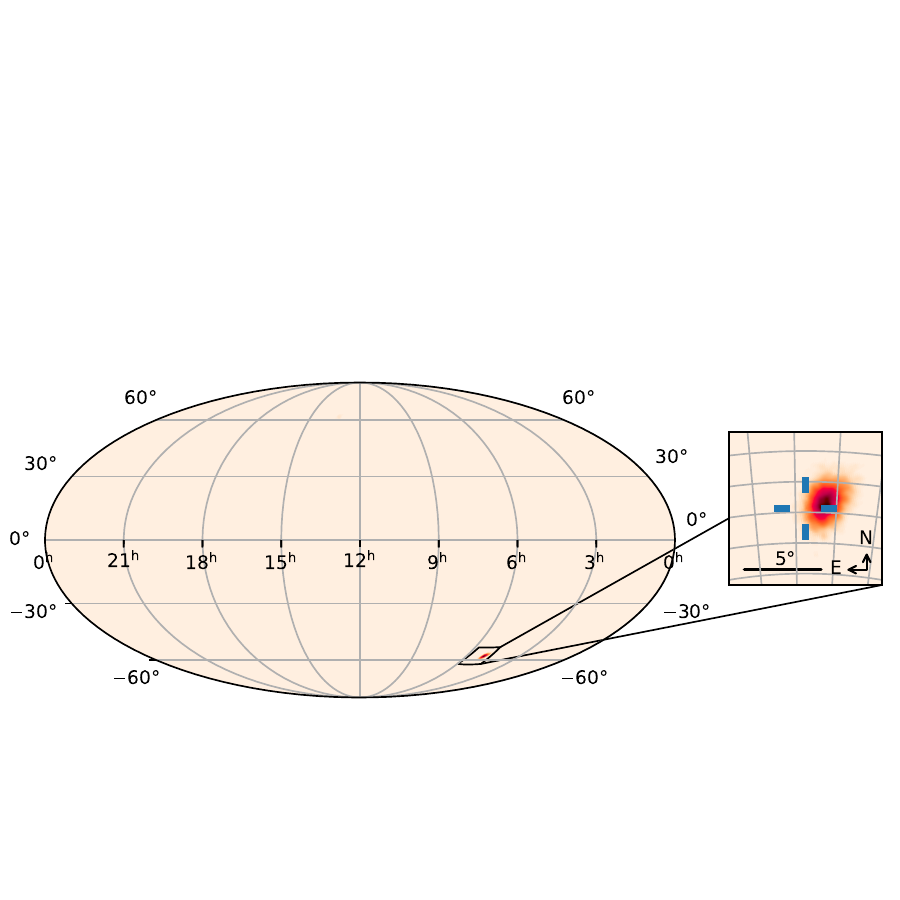}
    \caption{Example of analyses of a simulated signal with network SNR $\sim$50 that is observed in all three of the LIGO Hanford, LIGO Livingston, and Virgo detectors. The same parameters (besides the addition of Virgo) are used as in Fig.~\ref{fig:high_snr_2_det}. Due to the addition of Virgo, the signal is well-localized. Left: corner plot showing the posterior for the phasing parameters of the signal, including the phase-offset used in this work. Right: Inferred skymap for this simulated signal. Cross hairs in each panel indicate injected values. }
    \label{fig:high_snr_3_det}
\end{figure*}

In our study of the typical SNRs, we found the peaks of $\Delta \varphi$ are better constrained when three detectors are used in an analysis, as shown in Figs.~\ref{fig:high_snr_2_det} and \ref{fig:high_snr_3_det}. Physically, this is consistent with our expectations. The value of the phase $\varphi_{22}$ and the polarization angle $\psi$ are degenerate with one another, implying that the value of $\psi$ must be well-constrained in order to measure $\Delta \varphi$ well. In addition, $\psi$ is degenerate with the sky position parameters, $\alpha$ and $\delta$. If only the two LIGO detectors are used, without Virgo, the position of the signal cannot be well constrained, as is shown clearly in Fig.~\ref{fig:high_snr_2_det}, where the possible sky positions follow a large shape, corresponding to a poorly constrained $\Delta \varphi$. With Virgo's data, however, the sky position can be constrained into a single peak, as is shown in Fig.~\ref{fig:high_snr_3_det}, and this corresponds to a better constrained $\Delta \varphi$.  \emph{Thus, the use of three detectors is essential for constraining $\Delta \varphi$}, as it allows for the constraining of the sky localization parameters, $\psi$, and thus $\varphi_{22}$ and $\Delta \varphi$. 

\begin{figure*}[tb]
     \centering
    \includegraphics[width=0.44\textwidth]{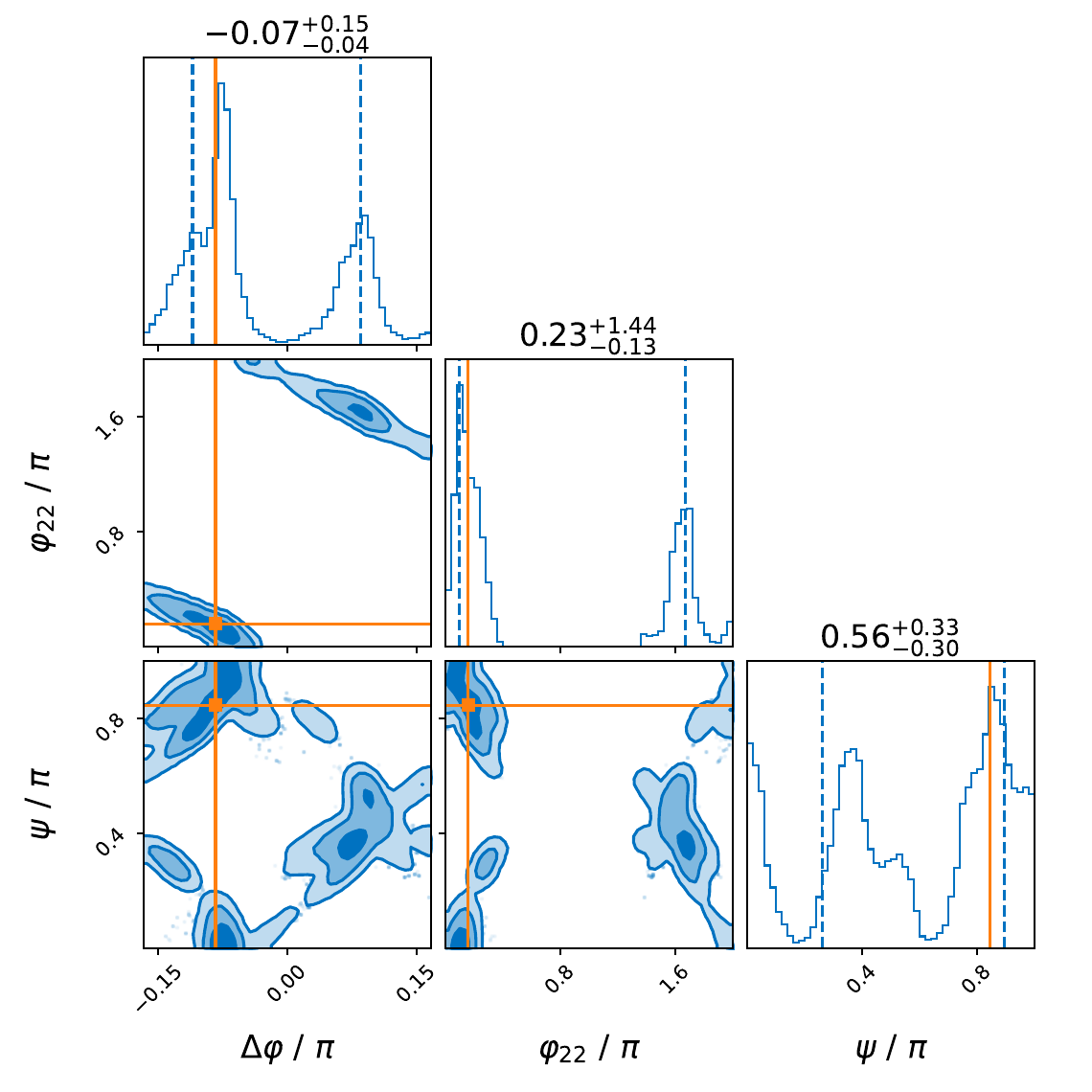}
     \includegraphics[trim={0 1cm 0 3cm},clip,width=0.54\textwidth]{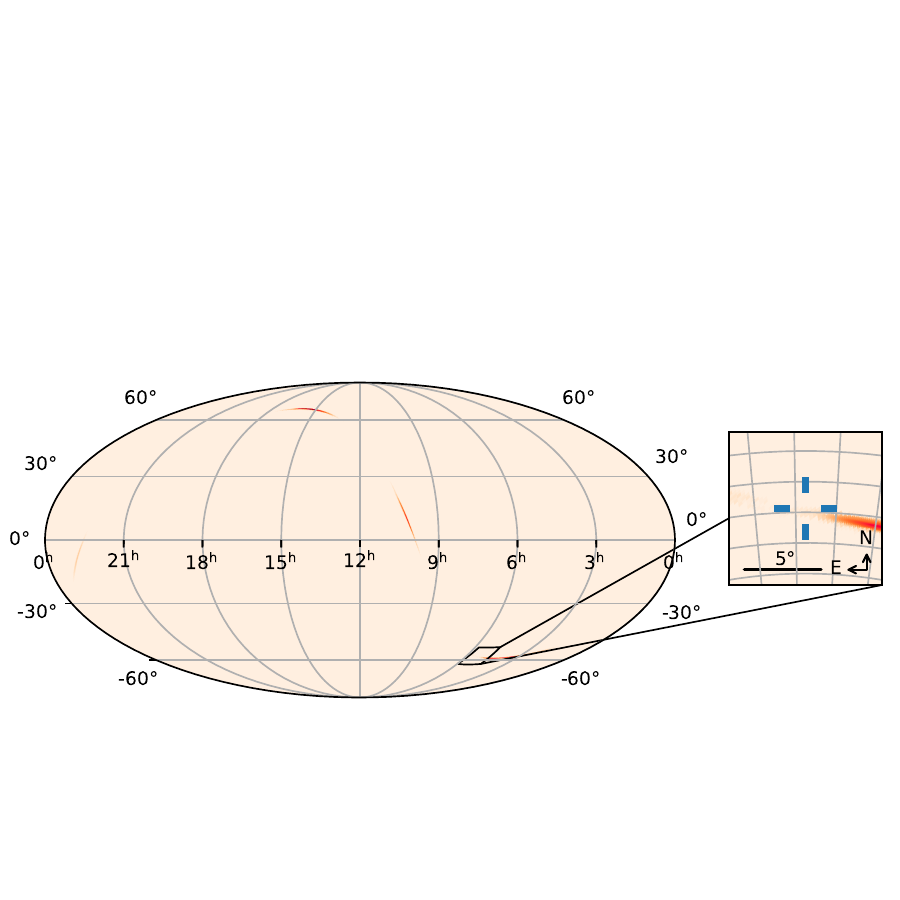} 
    \caption{Example of analyses of a simulated signal with network SNR $\sim$100 that is observed in the LIGO Hanford and LIGO Livingston detectors. Left: corner plot showing the posterior for the phasing parameters of the signal, including the phase-offset used in this work. Right: Inferred skymap for this simulated signal. Cross hairs in each panel indicate injected values. }
    \label{fig:vhigh_snr_2_det}
\end{figure*}

\begin{figure*}[tb]
    \centering
    \includegraphics[width=0.44\textwidth]{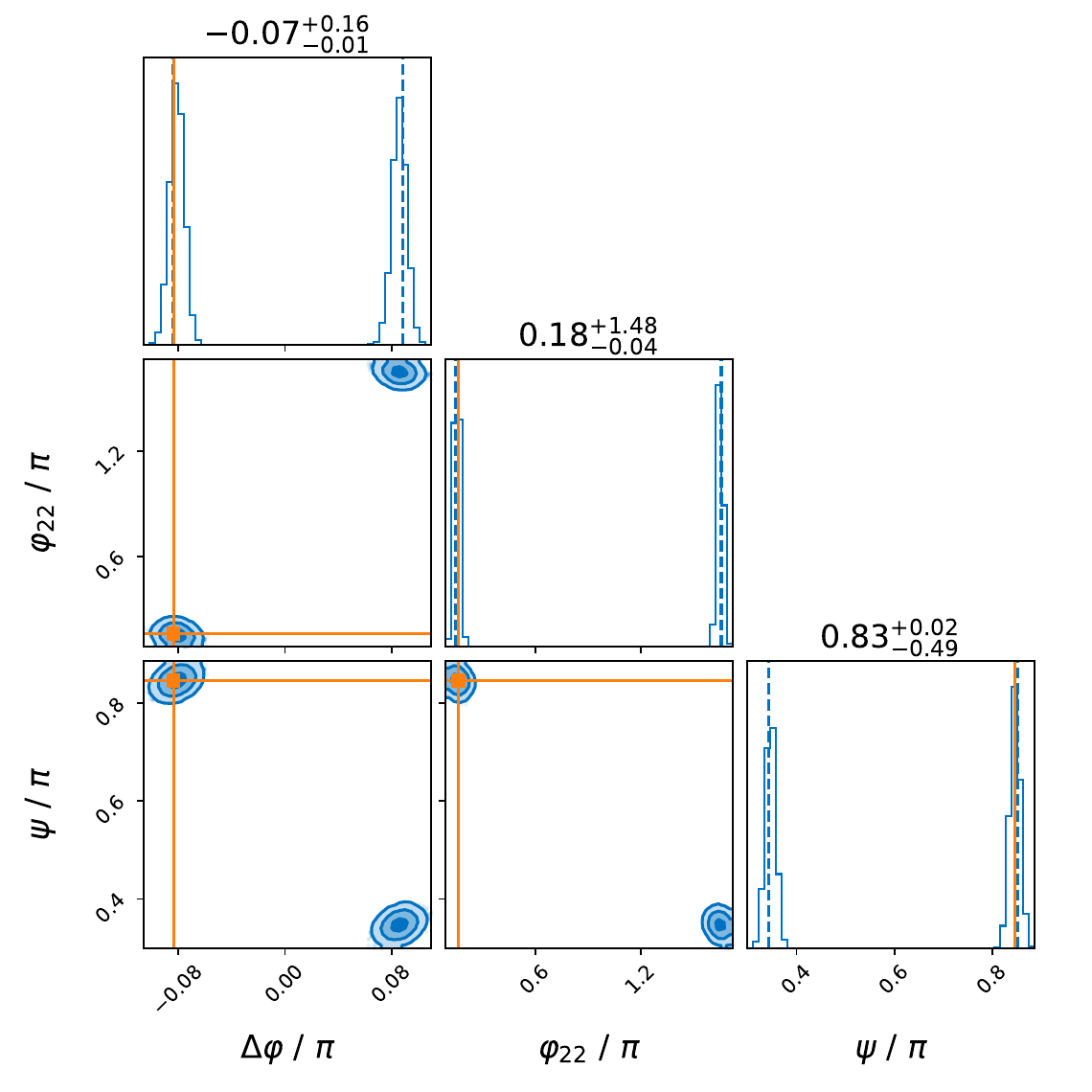}
    \includegraphics[trim={0 1cm 0 3cm},clip,width=0.54\textwidth]{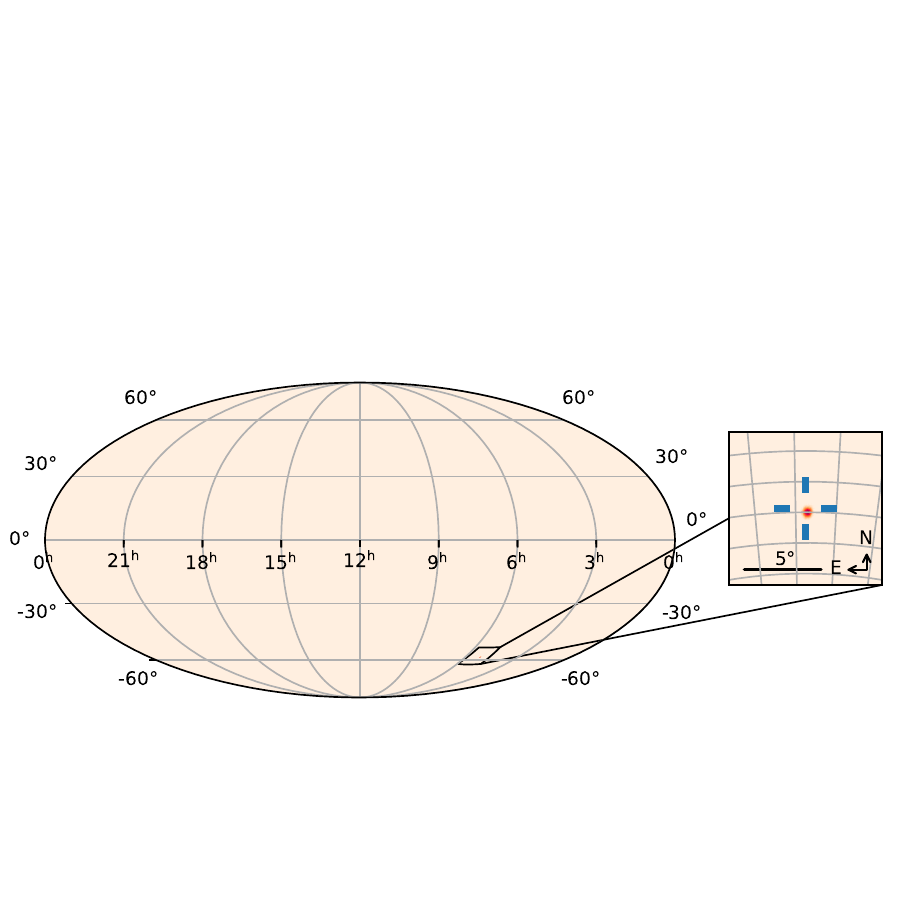}
    \caption{Example of analyses of a simulated signal with network SNR $\sim$100 that is observed in the LIGO Hanford, LIGO Livingston, and Virgo detectors. The same parameters (besides the addition of Virgo) are used as in Fig.~\ref{fig:vhigh_snr_2_det}. Due to the addition of Virgo, the signal is well-localized.  Left: corner plot showing the posterior for the phasing parameters of the signal, including the phase-offset used in this work. Right: Inferred skymap for this simulated signal. Cross hairs in each panel indicate injected values. }
    \label{fig:vhigh_snr_3_det}
\end{figure*}

In fact, even at extremely high SNRs, this pattern holds. In Fig.~\ref{fig:vhigh_snr_2_det}, we see that even at SNR 100, two detectors alone cannot determine the sky localization well enough to fully exclude $\Delta \varphi $ at 0, corresponding to a poorly constrained sky position. This contrasts with the three detector case in Fig.~\ref{fig:vhigh_snr_3_det}, in which we were able to produce clear peaks at $\Delta \varphi \approx \pm \pi/12,$ exactly as expected from the $\pi/6$ degeneracy in $\Delta \varphi$, corresponding to the clear peak in sky position. \emph{Thus, even if instrumental improvements allow for the detection of signals at higher SNR, the requirement of multiple detectors would hold.} Alternatively, if a next generation detector developed that could determine sky localization on its own, such as the proposed Einstein Telescope~\cite{ET_design}, such a detector could relax the need for multiple detectors to constrain the phase offset.

\subsection{Ratio of Posterior to Prior Accessible Range by SNR}

\begin{figure*}[t]
\includegraphics[width=\columnwidth]{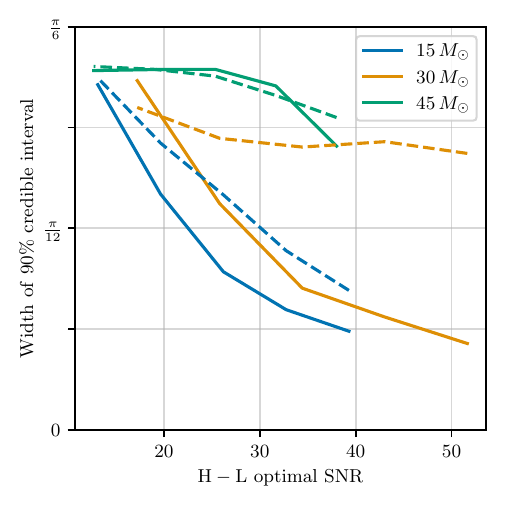}
\includegraphics[width=\columnwidth]{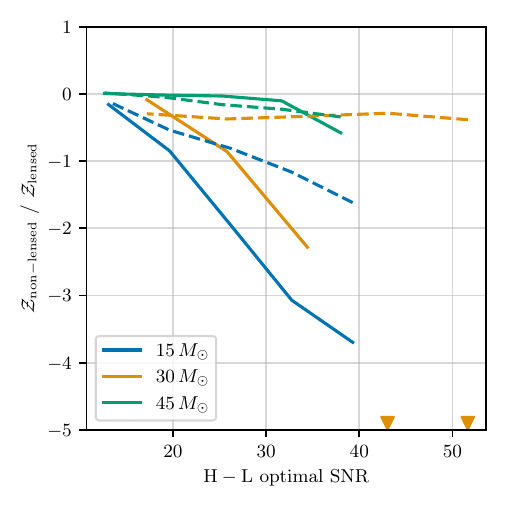}
\caption{\label{fig:sim_results} Results of signals injected at typical SNR values. \emph{Left:} the width of the 90\% credible interval for $\Delta \varphi$ for different chirp masses and SNR values. \emph{Right:} the non-lensed Bayes Factor (calculated using the Savage-Dickey ratio~\cite{dickey1970weighted}). Dotted lines represent results that only use the two LIGO detectors, while solid lines represent results that include Virgo in addition to the two LIGO detectors. Triangles  indicate values that are outside the presented axes. Results show that results at SNR $\geq 50$ and $\mathcal{M} \leq 30 \chirp$, should yield a reasonable constraint of $\Delta \varphi$ both away from zero phase shift and near the injected value.}

\end{figure*}

For signals injected at typical SNRs, the results are visualized in Fig.~\ref{fig:sim_results}, which depicts the percentage of the $\Delta \varphi$ prior covered by the 90\% credible interval. Results using two detectors are represented by dashed lines and those using three detectors are represented by solid lines. In general, we find results consistent with expectation, with higher SNRs and lower $\mathcal{M}$ leading to a better constraint of the phase offset. 
In addition, we can see the significant improvement the use of three detectors consistently has on specifying our results, as is most pronounced with the 50\% difference at SNR 50 for $\mathcal{M} = 30 \chirp$.

In general, we find that the signals are typically too quiet for SNRs at 30 or below to constrain $\Delta \varphi$ beyond the prior at a significant level. However, we find that at SNR 40, we can constrain the result to $25 \%$ of the prior for $\mathcal{M} = 15 \chirp$, with a similar constraint for $\mathcal{M} = 30 \chirp$ at SNR 50. \emph{This implies that we should be able to yield evidence of gravitational lensing for type II images at SNR $\gtrsim 40$, with $\mathcal{M} \leq 30 \chirp$} (assuming an asymmetric enough mass ratio so differences between the two modes are notable, the signal has enough precession to break the degeneracy between $\varphi_{22}$ and $\psi$, and the signal was measured with three detectors). This SNR and $\mathcal{M}$ limitation means that current GW signals are generally too quiet to yield a proper constraint of $\Delta \varphi$. However, as interferometer technology continues to advance, we expect to observe signals that can be reasonably tested for type II imagery using this method.  

We calculate the Savage-Dickey ratio~\cite{dickey1970weighted} to estimate the Bayes factor, $\mathcal{B}_{\Delta \varphi = 0}$ in support of the \emph{non-lensed} case, shown in the right panel of Fig.~\ref{fig:sim_results}. (Note that this is different from how a Bayes factor is canonically defined, where it is taken to be the ratio of the ``alternative hypothesis'', which does not occur frequently, over the ``null hypothesis''.) For the lensed hypothesis ($\mathcal{H}_{\Delta \varphi \neq 0}$) and the non-lensed hypothesis ($\mathcal{H}_{\Delta \varphi = 0}$), $\mathcal{B}_{\Delta \varphi = 0}$ is given by

\begin{equation}
\begin{split}
\mathcal{B}_{\Delta \varphi = 0} &= \frac{p(D|\mathcal{H}_{\Delta \varphi = 0})}{p(D|\mathcal{H}_{\Delta \varphi \neq 0})} \\
&= \lim_{\Delta \varphi\to 0} \left[ \frac{ p(\Delta \varphi|D,\mathcal{H}_{\Delta \varphi = 0}) }{ p(\Delta \varphi|\mathcal{H}_{\Delta \varphi = 0}) } \right]  \\
\end{split}
\end{equation}

 for some data $D$. Hence, $\mathcal{B}_{\Delta \varphi = 0}$ can be estimated by taking the ratio of the prior to the marginalized posterior at $\Delta \varphi = 0$.

We find that $\log_{10} \mathcal{B}_{\Delta \varphi = 0} <  -1$ against zero phase offset can be easily achieved for lower chirp mass BBH systems that are observed by three detectors. Specifically, we find strong support that rules out zero phase offset when $\text{SNR} > 25$ for $30 \chirp$ and $\text{SNR} > 20$ for $15 \chirp$. Consistent with Fig.~\ref{fig:sim_results}, we do not confidently identify a non-zero phase offset for the SNR range considered for systems of $45 \chirp$.

\section{Astrophysical Events} \label{sec:events}

Based on the previously described limitations of this test for type II images, 
we choose to further analyze two astrophysical events from~\cite{GWTC-3},
GW190412~\cite{GW190412} and GW190814~\cite{GW190814}. 
These two events are the most likely to have measurable phase differences between the (2,2) and (3,3) mode of the signal due to the high signal-to-noise ratio, asymmetric mass ratio, and small localization area for both events. 
Evidence for lensing was investigated for GW190412 in a previous LVK publication~\cite{GWTC-3_lensing}, 
but no investigations of GW190814 have been published to date. 
GW190814 is of particular interest for this study as it has the highest ``higher mode SNR'' for any event detected to date~\cite{Hoy:2021dqg} and has been claimed to be lensed~\cite{Broadhurst:2020cvm}.

\subsection{Parameter estimation results}

\begin{figure}[t]
\includegraphics[width=\columnwidth]{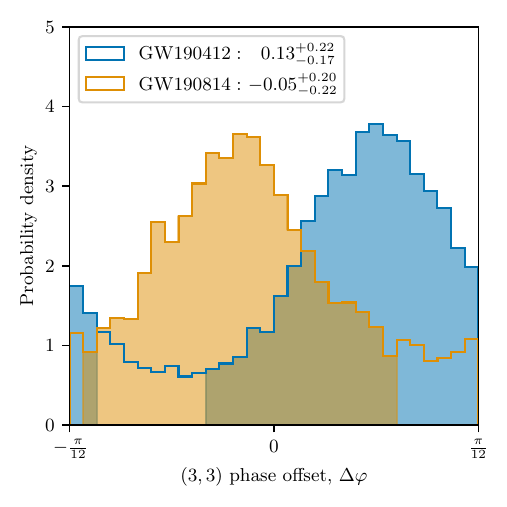}
\caption{\label{fig:real_results} Posteriors of the (3,3) offset, $\Delta \varphi$, for GW190412 and GW190814. The maximum a posteriori estimate of $\Delta\varphi$ is between 0 and $\pi/12$ for GW190412, but neither value is excluded at the 90\% credible level. The posterior for GW190814, on the other hand, peaks at $\Delta \varphi=0$.}
\end{figure}

We estimated the parameters of GW190412 and GW190814 using the same waveform model and settings as described for the simulation studies.
The posterior results for the (3,3) phase offset for both events can be seen in 
Fig.~\ref{fig:real_results}.
We measure $\Delta \varphi$ to be $0.13^{+0.22}_{-0.17}$ for GW190412 and $-0.05^{+0.20}_{-0.22}$ for GW190814.
We would expect $\Delta \varphi = \frac{\pi}{12} \approx 0.26$ in the case of a type II lensed signal.
We find that $\log_{10} \mathcal{B}_{\Delta \varphi = 0}$ to be $-0.14$ for GW190412 and $0.21$ for GW190814.
Hence, while $\Delta \varphi$ is marginally constrained in both cases, neither result is strong enough to confidently support the lack or presence of a phase offset. 
This result agrees with previous investigations of GW190412 that showed marginal support for being a type II lensed signal ~\cite{GWTC-3_lensing}. 
For GW190814, our results show marginal support in favor of zero phase offset and hence, not a type II lensed signal. 
\subsection{Expected uncertainties}

To investigate our phase offset results for these two astrophysical events further,
we can compare our results to the expected measurement uncertainty of $\Delta \varphi$ based on the match between a waveform with zero offset and nonzero offset.
We can use this to test a family of signals with properties consistent with GW190412 or GW190814, in order to find the luminosity distance threshold where the signals become loud enough to measure the phase offset. 
Given a true waveform $h(\theta_\text{true})$, the posterior can be estimated as a function of $\theta$ using the SNR of $h(\theta_\text{true})$, $\rho_\text{true}$, and the match between $h_\text{true}$ and some other waveform $h(\theta)$.
The match between these waveforms, $\epsilon(\theta)$ is
\begin{equation}\label{eq:match}
   \epsilon^2(\theta) = \frac{ \langle h(\theta_\text{true}) | h(\theta) \rangle^2 } { \langle h(\theta) | h(\theta) \rangle \langle h(\theta_\text{true}) | h(\theta_\text{true}) \rangle}, 
\end{equation}
where $\langle a|b \rangle$ denotes
\begin{equation}
   \langle a|b \rangle = 4 \left( \mathbb{R}\int_{0}^{\infty} \frac{a(f)b^*(f)}{S(f)} df \right)
\end{equation}
for a given power spectral density $S(f)$. The match is maximized over phase, which in this context can be considered as the maximum value of Eq.~\ref{eq:match} for any $\varphi_{22}$ with a given $\Delta \varphi$.  The posterior, $P(\theta | D)$, is then given by~\cite{LIGOScientific:2019hgc}
\begin{equation}
    P(\theta | D) \propto \exp \left[ -\frac{(1-\epsilon^2(\theta)) \rho_\text{true}^2}{2} \right] .
\end{equation}

\begin{figure*}[t]
\includegraphics[width=\columnwidth]{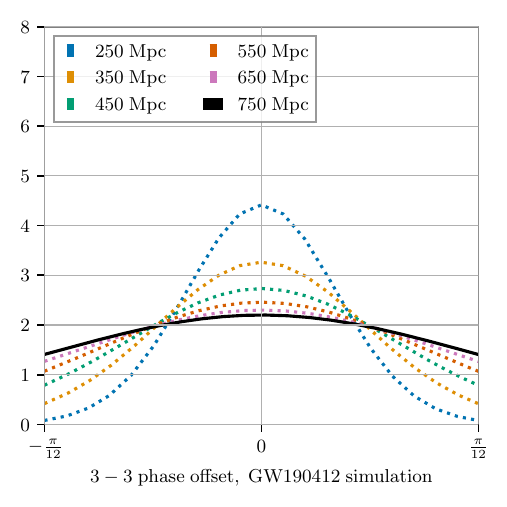}
\includegraphics[width=\columnwidth]{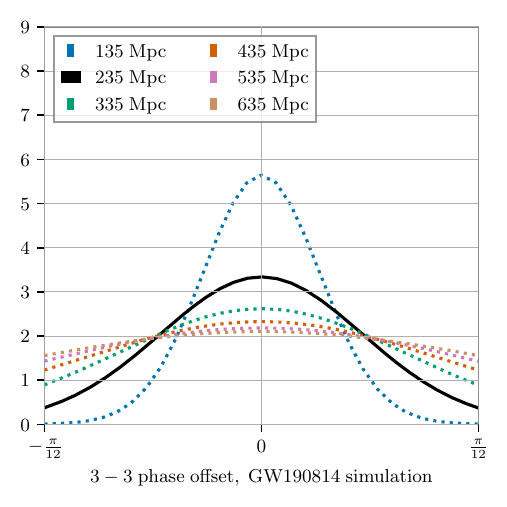}
\caption{\label{fig:real_estimates} Simulated posteriors based on match between waveforms at different distances for simulated events consistent with the properties of GW190412 and GW190814. Black lines indicate which simulated distance is most consistent with the measured properties of the real event. In the case of GW190412, these simulations suggest that the value of $\Delta \varphi$ is not expected to be well-constrained while $\Delta \varphi$ is expected to be weakly constrained. In both cases, closer (higher SNR) events in the future with similar properties are expected to constrain $\Delta \varphi$. }
\end{figure*}

Simulated posteriors generated with the above prescription for true waveforms with properties consistent with GW190412 and GW190814 at different luminosity distances can be seen in Fig.~\ref{fig:real_estimates}.
The estimated posterior for a signal at a distance consistent with GW190412 or GW190814 is shown as a black solid line, with other luminosity distances plotted with dotted lines. 

Based on Fig.~\ref{fig:real_estimates}, we find that our GW190412 measurement is more constraining than would be expected based on the measurability of the phase offset alone;
reasons for this discrepancy could include the effects of noise or additional correlations that we do not account for when generating simulated marginalized posteriors. 
Conversely, our measured uncertainty for GW190814 is consistent with expectations.  
In both cases, we find that these gravitational-wave events would have to have been at much lower luminosity distances for $\Delta \varphi$ to be confidently measured. 
These results qualitatively agree with the complete simulations shown in Fig ~\ref{fig:sim_results}.
While it will be possible to confidently identify or rule out type II lensed signals for sufficiently high SNR events, 
no event to date has been identified with sufficient SNR in the (3,3) mode to make an unambiguous identification of type II images.

\section{Comparison to previous results} \label{sec:compare}

Image type analysis of potentially lensed gravitational-wave signals has been previously investigated in a variety of situations, both by analyzing multiple images together or looking at single images~\cite{Lo:2021nae, Janquart:2021qov, Janquart:2021nus}. 
The analysis presented in this work differs most significantly from these works in how we choose to parameterize the measurable differences between different image types. 

As an example of this difference, we can compare this analysis to the recent investigation of GWTC-3 events~\cite{gwtc3_lensing} using the \texttt{GOLUM} pipeline~\cite{Janquart:2021qov, Janquart:2021nus}.
This analysis also investigated if there was support for phase differences between the different spherical harmonic modes but parameterized the offset using the Morse phase rather than directly using the phase difference between different spherical harmonic mode phases. 
However, the Morse phase $n_j = \left\{ 0, 1/2, 1\right\}$ (corresponding to type I, II and III, respectively \cite{1992grle.book}) is simply related to this phase offset, $\Delta \varphi$, by $\Delta \varphi = \pi n_j / 6$. 
In this way, the use of $\Delta \varphi$ can be treated as a reparameterization of $n_j$.

This parameterization is the most natural way to model this specific problem and addresses a significant issue with single-event analyses using Morse phase.
A significant benefit of using $\Delta \varphi$ is that it naturally allows the use of cyclical prior that accounts for the fact that 
type I and III images are indistinguishable for quasi-circular, non-precessing signals when considering only a single image~\cite{Ezquiaga:2020gdt}. 
If a cyclical prior is not used, the evidence for a type I vs III image could be non-zero despite the two scenarios being indistinguishable when the evidence was estimated from finite number of samples. 

An example of this aphysical behavior can be seen in the \texttt{GOLUM} results for GW190412~\cite{gwtc3_lensing}.
For this event, the authors note that $\log_{10} (\mathcal{B}^{II}_{I})=0.60 \pm 0.16$ (where $\mathcal{B}^{II}_{I}$ is the Bayes factor for the type II hypothesis versus the type I hypothesis), while $\log_{10} (\mathcal{B}^{II}_{III})=0.22 \pm 0.16$.
However, this also means $\log_{10} (\mathcal{B}^{I}_{III})\approx0.38 \neq 0$, despite type I and type III images being indistinguishable when considering only one image.

Conversely, when using $\Delta \varphi$, the indistinguishability of type I and III images in a single event is explicitly accounted for, as both scenarios would result in $\Delta \varphi = 0$. Using this parameterization when analyzing GW190412, we find that $\log_{10} (\mathcal{B}^{II}_{I})=\log_{10} (\mathcal{B}^{II}_{III})=0.13$, where $\mathcal{B}^{II}_{I}=\mathcal{B}^{II}_{III}$ by construction.
The choice of $\Delta \varphi$ is hence of more interest than the specific image type for a single-event analysis. 
For these reasons, $\Delta \varphi$ is the preferred parameterization for this analysis compared to $n_j$.

In addition, this analysis is similar to that presented in \emph{Capano and Nitz}~\cite{Spectroscopy}. In that paper, the authors allow multiple parameters to vary between the (2, 2) and (3, 3) modes, including the chirp mass, mass ratio, spins, and phase, with a focus on looking for effects not predicted by GR. In particular, their treatment of the phase offset $\Delta \varphi$ as the absolute difference between the two modes phase is the same as ours. In our case, however, we do not expect lensing to affect other parameters outside of the phase, and so we fix all non-phase parameters between the two modes. This allows us to focus on the relevant parameter for lensing, leading to better measurements.

Thus, looking for evidence of just the mode phase difference directly as a continuous parameter is advantageous, as this allows us to estimate the support for a type II lensed signal in the measurable parameter space. Furthermore, this can be generalized to generic phase shifts, allowing one to identify shifts not predicted by general relativity.

\section{Conclusions} \label{sec:conclusions}

We conclude that we have successfully developed a new search method for gravitationally lensed type II images in individual GW events by searching for a phase offset of $-\pi/12$ between the dominant (2, 2) mode and the higher order (3, 3) mode in signals. In particular, we have found that for a precessing signal with an asymmetric mass ratio detected by three detectors, we can recover the phase offset reasonably when $\mathcal{M} \leq 30 \chirp$ and H-L optimal SNR $ \geq 50$. We note as well the importance of constraining the sky location in this test, as at least three detectors must be used to reasonably constrain the phase offset. We note that more work can be done in fully characterizing the regime in which this test could feasibly detect gravitational lensing for type II images. For example, further work could examine how different mass ratios and spins affect the ability to detect the phase offset, or could examine offsets between even higher order modes ((2,2) and (4,4) for example) . 

Our results from analyzing GW190412 and GW190814 are in line with our expectations of the simulated analysis; neither result contains conclusive evidence for or against type II images, with GW190412 having only marginal support for such lensing and GW190814 having only marginal support against such lensing. However, the posteriors for the phase offsets agree qualitatively with the results of our simulated waveforms, with neither having high enough SNRs to constrain the phase offset.
We also find that the optimal scenario to evaluate the evidence of type II images is loud, well-localized events with low masses (assuming that the mass ratio is asymmetric enough to detect higher order modes).The loudest events observed with three or more detectors in upcoming observing runs are, hence, the best candidates to investigate for type II lensing. 

We thus highlight the importance of increasing the sensitivity of the gravitational wave detectors and increasing the number of gravitational wave detectors in looking for lensing. Both well-constrained sky location and higher signal SNRs are needed to probe individual events for type II imagery successfully.
These high SNR signals needed to identify lensing will become commonplace once the next generation of gravitational-wave detectors come online\cite{ET_design,CE_horizon}, allowing for the unequivocal identification of type II lensed signals.

\section*{Acknowledgements} \label{sec:acknow}
We thank Alan Weinstein for support and helpful discussions throughout this project. 
We thank Katerina Chatziioannou for discussions related to the expected phase behavior of Type II images and Lucy Thomas for her input on waveform choices and parameter estimation. 
We thank Mick Wright for helpful comments during internal review of this work. 
KT and DD acknowledge support from the NSF through the LIGO Laboratory. KT also acknowledges support from the US DoD through the NDSEG fellowship.
RKLL acknowledges support from the research grant no.~VIL37766 and no.~VIL53101 by the Villum Fonden, the DNRF Chair program grant no.~DNRF162 by the Danish National Research Foundation, the European Union's Horizon 2020 research and the innovation programme under the Marie Sklodowska-Curie grant agreement No.~101131233. The Center of Gravity is a Center of Excellence funded by the Danish National Research Foundation under grant No.~184.
This material is based upon work supported by NSF’s LIGO Laboratory 
which is a major facility fully funded by the 
National Science Foundation.
LIGO was constructed by the California Institute of Technology 
and Massachusetts Institute of Technology with funding from 
the National Science Foundation, 
and operates under cooperative agreement PHY-2309200.
The authors are grateful
for computational resources provided by the LIGO Laboratory and supported by National Science Foundation
Grants PHY-0757058 and PHY-0823459.

\section*{Data Availability} \label{sec:data}
The data relevant to our analysis can be found online \cite{taylor_2025_15786857}. 

\appendix

\section{Analysis details}~\label{app:details} 

In our analysis, we used several simulated waveforms, varying the chirp mass $\mathcal{M}$ to see how our results changed based on system masses and varying the luminosity distance $d_L$ in order to vary the signal SNR. In particular, we ran our trials with $\mathcal{M} = 15 \chirp$, $\mathcal{M} = 30 \chirp$, or $\mathcal{M} = 45 \chirp$. For our luminosity distances, we did two different sets of trials. The first consisted of signals at very high SNRs, with the same luminosity distances used for each $\mathcal{M}$ value, so that we could see general results without worrying about the constraints of quiet signals. The second consisted of signals with more typical SNRs roughly chosen to vary by 10 from 10 to 50, meaning different luminosity distances had to be chosen for different $\mathcal{M}$ values. The values of $d_L$ for each trial are given in Table \ref{tab:ld_Table}. 
\begin{table}[tb]
    \centering
    \begin{tabularx}{\columnwidth}{|X|X|}
    \hline
    Trial & Luminosity Distances\\
    \hline
    High SNR Trials (All Masses) & \{50, 60, 75, 100, 150, 300\} \\
    \hline
    Typical SNRs, $\mathcal{M} = 15$ & \{210, 252, 315, 420, 630\} \\
    \hline
    Typical SNRs, $\mathcal{M} = 30$ & \{290, 384, 435, 580, 870\} \\
    \hline
    Typical SNRs, $\mathcal{M} = 45$ & \{510, 612, 765, 1020, 1530\} \\
    \hline
    \end{tabularx}
    \caption{Luminosity Distance Parameters. The same set of luminosity distance was chosen for the high SNR trials, as we only cared about general results in this case. For the typical SNR trials, we chose different luminosity distances for each $\mathcal{M}$ to yield somewhat consistent SNR values between the three chirp masses  }
    \label{tab:ld_Table}
\end{table}

\begin{table}[tb]
    \centering
    \begin{tabularx}{\columnwidth}{|X|X|p{4cm}|}
    \hline
    Parameter & Type & Prior for Precession\\
    \hline
    $m_1$ & Constraint & $5$ M$_{\odot}$ $\leq m_1 \leq 60$  M$_{\odot}$\\
    \hline
    $m_2$ & Constraint &5 M$_{\odot}$ $\leq m_2 \leq60$  M$_{\odot}$\\
    \hline
    $\mathcal{M}$ & Uniform & $10$ M$_{\odot}$ $\leq \mathcal{M} \leq 60$  M$_{\odot}$\\
    \hline
    $q$ & Uniform & .1 $\leq q \leq 1$ \\
    \hline
    $a_{1}$ & Uniform & $0 \leq a_1  \leq .9$ \\
    \hline
    $a_{2}$ & Uniform & $0 \leq a_2 \leq .9$ \\
    \hline
    $\theta_{1}$ & Sine & $0 \leq \theta_1  \leq 2\pi$ radians\\
    \hline
    $\theta_2$ & Sine & $0 \leq \theta_2 \leq 2\pi$ radians\\
    \hline
    $\phi_{12}$ & Uniform  & $0 \leq \phi_{12}  \leq 2\pi$ radians\\
    \hline
    $\phi_{jl}$ & Uniform & $0 \leq \phi_{jl} \leq 2\pi$ radians \\
    \hline
    $d_L$ & Uniform (source frame) & 10 Mpc $\leq$ $d_L$  $\leq$  $2d_{L_{injected}}$\\
    \hline
    $\theta_{jN}$ & Sine & $0 \leq \theta_{jN}  \leq 2\pi$ radians\\
    \hline
    $\delta$ & Cosine & $0 \leq \delta  \leq 2\pi$ radians\\
    \hline
    $\alpha$ & Uniform & $0 \leq \alpha  \leq 2\pi$ radians\\
    \hline
    $\psi$ & Uniform & $0 \leq \psi  \leq \pi$ radians\\
    \hline
    $\varphi_{22}$ & Uniform & $0 \leq \varphi_{22} \leq 2\pi$ radians\\
    \hline
    $\Delta\varphi$ & Uniform & $-\pi/6 \leq \Delta \varphi \leq \pi/6$ radians\\
    \hline
\end{tabularx}
    \caption{Priors for the simulated trials. Most of these priors are standard astrophysical priors, with the $d_L$ and masses prior changed to account for high SNR signals and speed up trials.}
    \label{tab:priors_Table}
\end{table}

We kept all other parameters the same, choosing a precessing signal with an asymmetric mass ratio. We choose a mass ratio asymmetric enough for differences between the (2, 2) and (3, 3) mode to be notable~\cite{Mills:2020thr}. We included precession in our signal because of a degeneracy between the phase $\varphi_{22}$ and the polarization angle $\psi$. For non-precessing signals, this degeneracy is exact, leading to a less constrained $\Delta \varphi$. For precessing signals, the precession of the system modulates $\varphi_{22}$ and $\psi$ differently and the degeneracy is broken, allowing us to better constrain $\Delta \varphi$ compared to a fully aligned system.
However, it should be noted that non-zero precession is not required to measure $\Delta \varphi$.

In particular, we chose a mass ratio of $q = 0.25$. We also used moderate spin parameters, equal spin magnitudes with a slight precession, with parameters $a_1=0.5$, $\theta_1=0.5$ radians, $a_2=0.5$, $\theta_2=0.5$ radians,  $\phi_{12}=1.7$ radians, and $ \phi_{jl}=0.3$ radians. We chose arbitrary polarization and sky location parameters that are within typical astrophysical ranges: $\psi=2.659$ radians, $\theta_{jn}=3\pi/8$ radians, $\alpha=1.375$ radians, $\delta=-1.008$ radians. Additionally, our reference time of 1239082062 and reference phase $\varphi_{22}=0.5$ radians were chosen such that the detector noise we inject into is typical for the LIGO and Virgo detectors in O3. Lastly, our injected phase offset was $\Delta \varphi = -\pi/12$ radians, as this is the offset in type II signals. 

For the trials, we sample using the prior shown in Table \ref{tab:priors_Table}. We used typical astrophysical priors for the mass ratios, spins, polarization parameters, right ascension, declination, and phase. We restricted our prior for our luminosity distance to be within twice the injected distance and our prior for the masses to have a lower upper limit, as this would allow our simulations to run faster without changing the sampling results. In addition, our new parameter $\Delta \varphi$ needed only to be sampled from $-\pi/6$ to $\pi/6$, due to the effects of the $\pi/6$ degeneracy.

Additionally, for all trials, we sample each signal twice: once using only data from the LIGO Hanford and Livingston detectors, and once using data from LIGO Hanford, LIGO Livingston, and Virgo detectors. This allows us to note the number of detectors needed to reasonably specify the value of $\Delta \varphi$.

\bibliography{bib.bib}

\end{document}